\documentclass[english]{article}

\usepackage{caption}
\usepackage{hyperref}
\usepackage[a4paper, total={6in, 8.1in}]{geometry}
\usepackage[ruled]{algorithm2e}
\usepackage{amsmath}
\usepackage{color}
\usepackage{enumitem}
\usepackage[pdftex]{graphicx}
\usepackage{amsfonts}
\usepackage{authblk}

\renewcommand{\algorithmcfname}{ALGORITHM}
\SetAlFnt{\small}
\SetAlCapFnt{\small}
\SetAlCapNameFnt{\small}
\SetAlCapHSkip{0pt}
\IncMargin{-\parindent}
\newtheorem{remark}{Remark}
\newtheorem{definition}{Definition}
\newtheorem{theorem}{Theorem}

\begin{document}


\title{NeatSort - A practical adaptive algorithm}


\author[1]{Marcello La Rocca}
\author[2]{Domenico Cantone}
\affil[1]{Scuola Superiore Sant'Anna}
\affil[2]{Universit\`a di Catania}

\date{}
\maketitle

\begin{abstract}
We present a new adaptive sorting algorithm which is optimal for most
disorder metrics and, more important, has a simple and quick
implementation.  On input $X$, our algorithm has a theoretical $\Omega
(|X|)$ lower bound and a $\mathcal{O}(|X|\log|X|)$ upper bound,
exhibiting amazing adaptive properties which makes it run closer to
its lower bound as disorder (computed on different metrics)
diminishes.  From a practical point of view, \textit{NeatSort} has
proven itself competitive with (and often better than) \textit{qsort}
and any \textit{Random Quicksort} implementation, even on random
arrays.
\end{abstract}








\section{Introduction}
\label{introduction}

Our algorithm \textit{NeatSort} is based on a simple idea: exploit all
the information one gathers while reading the input array, as soon as
one gets it.  It is in this good practice that \textit{NeatSort} ``cleverness"
resides.  \textit{NeatSort} is a variant of the standard
\textit{Mergesort} algorithm, as its core workflow consists of
merging ordered lists.  However, in order to speed up the merging phase, the input array $X$ is preliminarily scanned so as to split it into a (minimal) sequence of nondecreasing sublists $L[0],L[1],\ldots,L[m]$, by executing the following instructions:
\begin{enumerate}
\setcounter{enumi}{-1}
  \item $i=0$;
  
  \item add the first undiscovered element, $X[i]$, to a new sublist;
  
  \item keep adding elements $X[i+1],\ldots,X[k]$ to the current sublist
  until either $ X[k] > X[k+1] $ or $ k = |X| $;
      
  \item if there are still undiscovered elements in $X$, go back to
  step 1.
\end{enumerate}

The following properties are immediate:
\begin{enumerate}[label=(\Alph*)]
  \item\label{wasOne} each sublist $ L[q] $ is in nondecreasing order, for $ q = 0,1, \ldots, m $; 
  
  \item\label{wasTwo} if $ L[q][i_q] $ is the last element in $ L[q] $, then $L[q+1][0] < L[q][i_q]$, for $q=0,1, \ldots, m-1$;
  
  \item\label{wasThree} let $ L^*[q] $ and $ L^{*}[q+1] $ be, respectively, a sorted list resulting from merging $L[q]$ with any subset of the lists $L[0],\ldots,L[q-1]$, and a sorted list resulting from merging $L[q+1]$ with any subset of the lists $L[q+2],\ldots,L[m]$, where $q
  \in \{1,\ldots,m-1\}$.  Then $ L^*[q+1][0] < L^*[q][i^*_q] $, where 
  $i^*_q$ is the index of the last element in $ L^*[q] $.
\end{enumerate}

After creating the sublists $L[q]$, for $q=0,1,\ldots,m$, adjacent pairs can be merged using an \emph{ad hoc} variant of \textit{mergesort}'s merging procedure (which takes advantage of properties \ref{wasTwo} and \ref{wasThree} above), until a single (ordered) list remains.

In fact, property \ref{wasTwo} allows one to save one comparison when merging the initial lists, and then, thanks to property \ref{wasThree}, one can take advantage of such saving at each subsequent merging step of ``superlists".

%
%
%

\subsection{Merging points}
\label{merging_points}

Adjacent sublists $L[q],L[q+1]$, where as above $L[q]$ and $L[q+1]$ are in nondecreasing order and $L[q+1][0] < L[q][i_q]$ holds (with $i_q$ the index of the last element in $L[q]$), can be stably merged into a single nondecreasing list in a convenient way. For the sake of simplicity, let us first assume that 
\begin{equation}
\label{eq_condition}
L[q][0] < L[q+1][0] \quad \text{and} \quad 
L[q][i_q] > L[q+1][i_{q+1}] 
\end{equation}
hold.
Then, in order to merge $L[q]$ and $L[q+1]$, it is enough to find out two sequences 
\begin{equation}
\label{eq_mergingPoints}
0 < j_0 < j_1 < \ldots < j_t = i_q +1 \quad \text{and} \quad 
0 = k_0 < k_1 < \ldots < k_t = i_{q+1} +1
\end{equation}
of \emph{merging points} in $L[q]$ and $L[q+1]$, respectively, such that
\begin{alignat}{3}
L[q][j_i-1] &\leq L[q+1][k_i] &<& L[q][j_i] \label{firstRec}\\
L[q+1][k_{i+1}-1] &< L[q][j_i] &\leq& L[q+1][k_{i+1}]\label{secondRec}
\end{alignat}
%
for $i=0,1,\ldots,t-1$ (where we convene that $L[q+1][i_{q+1} +1] = +\infty$).
Then the array resulting from concatenating the slices\footnote{For an array $T$ of length $n$ and indices $0\leq i \leq j \leq n-1$, we denote by $T[i\,..\,j]$ the \emph{slice} of $T$ from $T[i]$ to $T[j]$. When $j < i$, $T[i\,..\,j]$ will denote the empty array.}
\[
L[q][0 \,..\,j_0],~
L[q+1][k_0 \,..\,k_1-1],~
L[q][j_1 \,..\,j_2-1],~
\ldots,~
L[q+1][k_{t-1} \,..\,k_t-1],~
L[q][j_{t-1} \,..\,j_t-1]
\]
(in the order shown) is the stable merging of $L[q]$ and $L[q+1]$.

\begin{remark}
By relaxing (\ref{eq_mergingPoints}) so as to allow $0 \leq j_0$ and $j_{t-1} \leq j_t$, the above considerations can be immediately generalized also to the cases in which any of the conditions in (\ref{eq_condition}) does not hold.
\end{remark}

The merging points $j_0,j_1,\ldots,j_t$ and $k_0,k_1,\ldots,k_t$ can be computed quite efficiently. The index $j_0$ can be found by performing a binary search in $L[q][0\,..\,i_q-1]$, as it is known in advance that $L[q][i_q] > L[q+1][0]$. Then, the remaining merging points can be found by a simple linear search which is directly based on the very definitions (\ref{firstRec}) and (\ref{secondRec}). The number of comparisons for two lists of length $k_q$ and $k_{q+1}$ is at most $O(\log(k_q) + (k_q + k_{q+1})) = O(k_q + k_{q+1})$; in every step the number of comparisons is therefore $O(|X|)$: it is self evident in the last step of the merging phase, with just two sublists with a total of $|X|$ elements to merge, but of course in every merging step the sum of the number of elements of all the sublists is always equal to $|X|$. Despite the asymmetry in the sublists sizes that is due to the very nature of the analysis phase, their number is guaranteed to be at most $\left\lceil \frac{|X|}{2} \right\rceil$, and at each merging step the number of sublists is halved, so there will be at most $O(\log|X|)$ merging steps, and therefore the total number of comparison is guaranteed to be $O(|X|\log|X|)$.

\subsection{Keys to improvements}

The standard \textit{mergesort} algorithm follows a strategy divided into two phases:
\begin{itemize}
  \item A top-down phase, where the initial array is recursively divided in half-sized subarrays, until a minimum size (1 element) is reached.
\item A bottom-up phase, where the subarrays are recursively merged back together, resulting in the (stably) sorted version of the initial array.
\end{itemize}

\begin{figure}[Ht]
\includegraphics[width=0.9\textwidth]{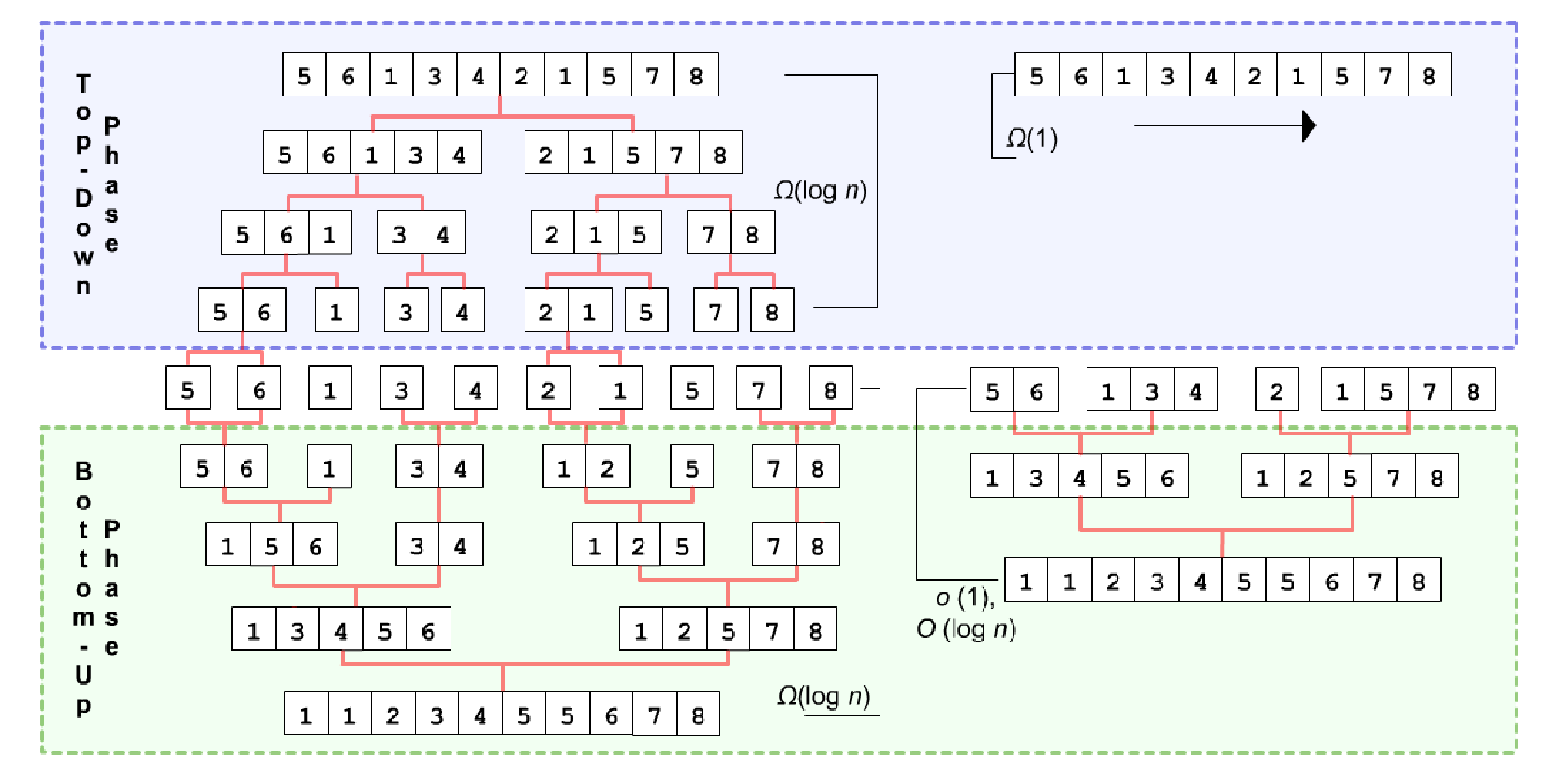}
\caption{A comparison between \textit{Mergesort} and \textit{NeatSort}}
\label{fig:one}
\end{figure}

In \textit{NeatSort}, the top-down phase is replaced by the preliminary phase which identifies the sequence of ordered sublists, as seen above. The latter will be used as the base for a subsequent bottom-up phase, which, up to some optimizations, is basically the same as in the standard \textit{Mergesort}.
Figure~\ref{fig:one} shows the different ways in which \textit{Mergesort} and \textit{NeatSort} work.

It is important to notice that, for an input array $X$, \textit{Mergesort}'s bottom-up phase requires $2\cdot|X|$ steps, whereas \textit{NeatSort}'s preliminary phase requires just $|X| - 1$ comparisons.


\section{Further improvements}
\label{sec:further_improvements}
The crucial improvement in \textit{NeatSort} is the efficient partitioning of the initial array in sublists during the preliminary phase, before the bottom-up phase starts. Notice, however, that in the worst case, i.e., when the initial array is sorted backwards,  $|X|$ sublists (containing exactly one element each) would be produced, thus resulting in no improvements in comparison to \textit{Mergesort}.

For the sake of clarity, let us suppose that our initial array $X$ is in strictly decreasing order, while a nondecreasing order is seeked for.
A first immediate solution would be to check, at the end of the preliminary phase, whether the number of sublists produced is greater than or equal to $\left\lceil\frac{|X|}{2}\right\rceil$: this could happen if and only if the ratio of adjacent elements which are inverted is higher than 50\%; in this case, the preliminary phase could just be repeated by examining the input array backwards (we denote it as \textit{backward analysis}, as opposed to \textit{forward analysis}, where array's elements are examined from first to last), and be sure to obtain an improvement. 

Settling with this solution, however, would betray \textit{NeatSort}'s philosophy of making use of all of the information one has collected.
Additionally, such solution is not optimal. In fact, let us consider the following array $X$, where
\begin{itemize}
  \item the first half contains $\left\lfloor\frac{|X|}{2}\right\rfloor$ elements in increasing ordered,
  \item the second half contains $\left\lceil\frac{|X|}{2}\right\rceil$ elements in decreasing order.
\end{itemize}

The analysis phase would produce a partitioning consisting of one list in account of the first half, plus $\left\lceil\frac{|X|}{2}\right\rceil$ lists in account of the second half, so that the backward analysis would take place and output one list for the second half of the 
initial array plus $\left\lfloor\frac{|X|}{2}\right\rfloor$ lists for the  first half of the input array, for a total number of lists equal to $\left\lfloor\frac{|X|}{2}\right\rfloor +1$.
This would be inefficient, as we know that the first half of the array is ordered, and so it is the second one (though in nonincreasing order). Thus, if the order of the second half is reversed, one ends up with just \emph{two} lists, rather than $\left\lfloor\frac{|X|}{2}\right\rfloor +1$ lists.

A solution to the above situation is the following: every time, during the preliminary phase, a singleton sublist is created (i.e., there is an inversion in the input, whose first element is not part of any previously created sublist), a new sublist formed by such two elements is created and then further elements are added to it until one is found which is greater than its predecessor--basically, a backward analysis is started from the point of the inversion to the first non-inverted couple of adjacent elements; subsequently, 
the sublist so obtained is reversed and a check is made to see if any additional element can be added to its tail (by any means starting a new forward analysis).

In the particular situation in which the input is sorted in reverse order, the above procedure creates just one list, proving itself as efficient as it is when dealing with sorted arrays (i.e., it is optimal in both extreme situations).

In the situation reported above, when the array is composed by two subarrays--the first one in increasing order and the second one in decreasing order,--such solution would create, during the preliminary phase, two lists; in particular, the construction of the second list would require $\left\lfloor \frac{\left\lceil \frac{|X|}{2}\right\rceil}{2}\right\rfloor$ element swaps (the first element in the sublist is swapped with the last one, the second one with the second-last one, etc.), and thus a total of $3 \cdot \left\lfloor \frac{\left\lceil \frac{|X|}{2}\right\rceil}{2}\right\rfloor$ assignments would be required.

After the preliminary phase, adjacent sublists are iteratively merged together 
using an \emph{ad hoc} variant of the canonical merge procedure until a single list is left.

\subsection{Correctness}

Let $L[0], L[1], \ldots, L[m]$ be the sequence of sublists created during the preliminary phase (with forward and backward analyses). Then, it is an easy matter to check that, by the very construction, the following two properties hold:
\begin{enumerate}[label=(\Alph*)]
  \item each sublist $ L[q] $ is in nondecreasing order, for $ q = 0,1, \ldots, m $; 
  
  \item if $ L[q][i_q] $ is the last element in $ L[q] $, then $L[q+1][0] < L[q][i_q]$, for $q=0,1, \ldots, m-1$.
\end{enumerate}

Properties \ref{wasOne} and \ref{wasTwo} readily imply

\begin{enumerate}[label=(\Alph*),start=3]
  \item let $ L^*[q] $ and $ L^{*}[q+1] $ be, respectively, a sorted list resulting from merging $L[q]$ with any subset of the lists $L[0],\ldots,L[q-1]$, and a sorted list resulting from merging $L[q+1]$ with any subset of the lists $L[q+2],\ldots,L[m]$, where $q
  \in \{1,\ldots,m-1\}$.  Then $ L^*[q+1][0] < L^*[q][i^*_q] $, where 
  $i^*_q$ is the index of the last element in $ L^*[q] $.
\end{enumerate}

From Property \ref{wasThree}, it follows that during any sequence of merging steps, in which only adjacent sublists are allowed to be merged, Properties \ref{wasOne} and \ref{wasTwo} are maintained as invariant, and so also Property \ref{wasThree}.

\subsection{Analysis phase performance}

The combination of forward and backward analysis proves itself optimal in any other situation with respect to the number of sublists created.

\begin{figure}
\centerline{\includegraphics[width=120mm]{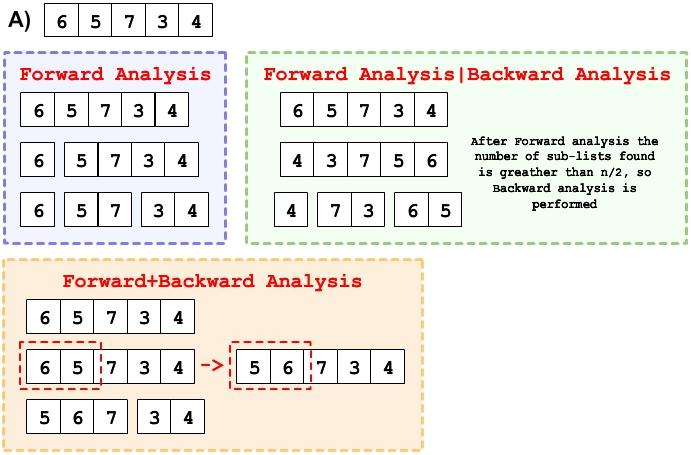}}
\centerline{\includegraphics[width=120mm]{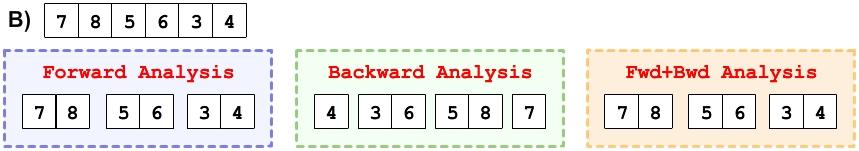}}
\centerline{\includegraphics[width=120mm]{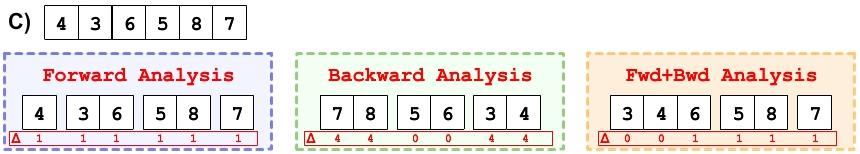}}
\centerline{\includegraphics[width=120mm]{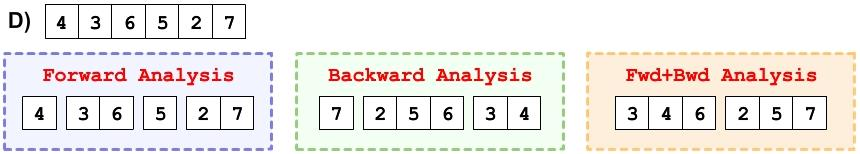}}
\centerline{\includegraphics[width=120mm]{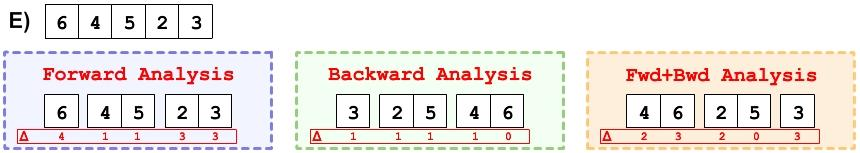}}
\centerline{\includegraphics[width=120mm]{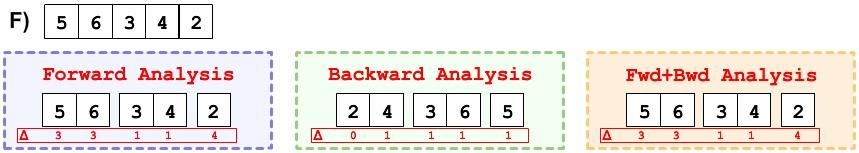}}
\caption{Performance of the different strategies described for the preliminary phase over a few examples. $\Delta$ indicates the distance of each element from its position in the sorted sequence.}
\label{fig:two}
\end{figure}

We can compare forward analysis, backward analysis and their combination through some examples shown in Figure~\ref{fig:two}.

As it is clear in each one of the examples above, the combination of forward and backward analysis produces a minimal number of sublists in comparison with:
\begin{enumerate}
  \item Mergesort (which will produce exactly $|X|$ sublists);
  \item Forward analysis only (by definition)
  \item The algorithm that applies forward analysis and than, if the number of sublists produced is greater than $\left\lceil \frac{|X|}{2} \right\rceil$, switches to backward analysis.
\end{enumerate}
While the correctness of each one of the statements above appears evident, for the first two the proof is trivial, while proving the last one, though intuitive, involves a simple reasoning by contradiction, which is left to the reader.

This solution, however, is not always optimal with respect to the total distance (the sum of the distances of each element from its final position in the ordered sequence), as can be seen in examples E and F, where backward analysis produces the lowest value; example C, however, shows how backward analysis can also lead to the highest possible value in other situations, so that backward analysis doesn't prove optimal either with respect to total distance.

\subsection{Heuristics}

In order to improve merging efficiency, a few attempts have been made.
First, as described in Section~\ref{merging_points}, different strategies have been tried to improve efficiency in finding merging points between lists and to improve the process.
\newline
Let $ L = \langle l_1, l_2, \ldots, l_n \rangle $ and $ R = \langle r_1, r_2, \ldots, r_m \rangle $  be two sublists in nondecreasing to be merged such that
$l_n > r_1$, and let $S = \langle s_1, s_2, \ldots, s_{n+m} \rangle $ be the list resulting from their merge.
Due to the nature of the problem and the overhead introduced to make an extra copy of at least one of the lists, the best performance has been reached with the procedure outlined in Algorithm~\ref{alg:one}.
\begin{algorithm}[t]
\SetAlgoNoLine
\KwIn{two sublists $ L = \langle l_1, l_2, \ldots, l_n \rangle $ and $ R = \langle r_1, r_2, \ldots, r_m \rangle $ in nondecreasing order and such that $l_n > r_1$.}
\KwOut{A single ordered list $S$ containing all the elements in the input lists.}

\medskip

Using binary search, find the lowest element $l_{i_1}$ in $ \langle l_1, l_2, \ldots, l_{n-1} \rangle $ greater than $r_1$ (i.e., the final position of $r_1$ in $S$). 

\tcp{Note that $1 \leq i_1 \leq n$. 
Thus the first $i_1$ elements in $S$ will be $ \langle l_1, l_2, \ldots, l_{i_1-1}, r_1 \rangle $, where, if $i_1 = 1$, the initial sublist $ \langle l_1, l_2, \ldots, l_{i_1-1} \rangle $ is empty.}

Init\footnotemark  $S = \langle l_1, l_2, \ldots, l_{i_1-1} \rangle $ and $ T = \langle l_{i_1}, \ldots, l_n \rangle $;

$k$ := $1$;

$i_1$ := $1$;

$j_1$ := $1$;

\Repeat{either $T$ or $R$ is empty}{
  
  
  
  add $r_{j_k}$ to the tail of $S$;

  $j$ := $1$;
  
  \Repeat{$r_{j_k + j} > t_{i_k}$ or $R$ is empty}{
    add $r_{j_k + j}$ to the tail of $S$;
    
    $j$ +:= $1$;
  }
  $j_{k+1} :=  j_k + j$;
  
  add $t_{i_k}$ to the tail of $S$;
  
  $i$ := $1$;
  
  \Repeat{$t_{i_k + i} > r_{j_{k+1}}$ or $T$ is empty}{
    add $t_{i_k+i}$ to the tail of $S$;
    
    $i$ +:= $1$;
  }
  $i_{k+1} := i_k + i$;
  
  $k$ +:= $1$;
}
\If{$R$ is empty}{
  copy all the elements left in $T$ to the tail of $S$;
}
\KwRet{$S$;}
\caption{NeatMerge}
\label{alg:one}
\end{algorithm}

\footnotetext{In the actual implementation, in order to minimize the number of swaps and extra memory consumption, S reuses the array L while T, that is a temporary array, will have its element copied from L[$i_1$], \ldots, L[n-1]; initially the size of array S is set to $i_1-1$ (possibly 0), and it will grow to $n+m$ elements, reusing the memory previously occupied by both L and R.}

Particular care has also been put in tuning the code.
In order to further improve performance, our efforts have been focused on the choice of the order used to merge the sublists: to introduce adaptivity in the merging phase (then having a second-level adaptivity), a few heuristics have been tested and compared against the simplest \textit{merge} approach, to verify whether possible advantages deriving from the choice of a better order for merging would be larger than the required overhead. Notice that the number of merges for merging $m$ lists is $(m-1)$, independently of the strategy followed.
\newline
We have tested and benchmarked the following alternative solutions:

\begin{enumerate}[label=\textbf{(\Roman*)}]
  \item \textbf{The first (leftmost) list is always merged with the second one.} 
  
This heuristic is sensibly slower than merging adjacent pairs. These results showed us that there might be a close connection between the degree of similarity between the size of the lists to be merged and the performance of the algorithm, which in turn suggested us to try to improve the coupling of the sublists in order to have their sizes matching as much as possible. The slowdown registered when merging unbalanced lists is likely to be related to the ratio of elements of the bigger list that has to be moved for each element in the smallest one: the largest the difference, the highest the ratio, until this turns into a bottleneck.

\item When one chooses to merge all pairs of adjacent lists, when the number of lists is odd, one of the lists go unaltered to the next step; usually, the surviving list is the last (rightmost) one. However, a 2\% improvement in execution time has been observed by \textbf{choosing to leave out the longest one}.

\item For a triple $A$, $B$, $C$ of adjacent lists (where $A$ precedes $B$ and $B$ precedes $C$), one checks whether $|A| \geq p(|B| + |C)$ holds, for an assigned constant $p$. If this is the case, lists $B$ and $C$ are merged whereas $A$ goes unaltered to the next step, otherwise $A$ and $B$ are merged.

A series of tests has been run to tune the parameter $p$; experimental results show that the best performance is obtained for values of $p$ ranging from $1.4$ to $1.25$, as the size of initial arrays grows from a few hundreds to millions of elements.
Using an average value for $p$, we obtained a performance improvement close to 3.2\%. The pseudocode of the resulting algorithm is shown in the box for Algorithm~\ref{alg:two}.
\end{enumerate}

%

\begin{algorithm}[H]
\SetAlgoNoLine
\KwIn{A list $X$.}
\KwOut{ The ordered version of the input list.}
$lists$ := [];
$listCounter$ := $1$;
add $X[1]$ to $lists[listCounter]$;


\For{$i$ := $1$ to $|X|$}{
	\While {$ X[i] \leq X[i+1] $}{
		add $X[i]$ to $lists[listCounter]$
		$i$ +:= $1$;
	}
	\If{$length(lists[listCounter])$ == $1$}{
		\While{$X[i] > X[i+1]$}{
			\textbf{append} $X[i+1]$ to $lists[listCounter]$;
			
			$i$ +:= $1$;
		}
		\textbf{reverse} $lists[listCounter]$;
		
		
		\While{$X[i] \leq X[i+1]$}{
			\textbf{append} $X[i+1]$ to $lists[listCounter]$;
			
			$i$ +:= $1$;
		}
  }
  \If{$listCounter > 1$ and first element in $lists[listCounter]$ is greater than or equal to the last element in $lists[listCounter-1]$}{
			\textbf{merge} $lists[listCounter-1]$ and $lists[listCounter]$
  }
	
	$listCounter$ +:= 1;
}
	
\While{$listCounter > 1$}{
	$j$ := $1$;
	
	\While{$j < listCounter$}{
		\If{$length(lists[j]) \leq p * ( length(lists[j+1]) + length(lists[j+2]) )$}{
			\textrm{neatMerge}$(lists[j], lists[j+1])$;
			
			$j$ +:= $2$;
		}
		\Else{
			\textrm{neatMerge}$(lists[j+1],lists[j+2])$;
			
			$j$ +:= $3$;
		}
  }
	$listCounter$ := $|lists|$;
} 
\caption{NeatSort}
\label{alg:two}
\end{algorithm}

\subsection{Asymptotic Analysis}
\label{sec:asymptotic_analysis}
Upper and lower bounds for \textit{NeatSort} can be computed quite trivially.
Given an array $X$ of length $n$, the preliminary phase requires $\Theta(n)$ time, while the merging phase, as in \textit{Mergesort}, requires $\mathcal{O}(n\log n)$ time: thus, the total time required by \textit{NeatSort} is $mathcal{O}(n\log n)$.
\newline
As for space requirements, the preliminary phase can be realized efficiently with an array of length at most $|\frac{n}{2}|$, while the merging procedure requires an array of length at most $n$, so the additional space required is $\mathcal{O}(n)$.
\newline
Summing up, denoting with $T(n)$ and $S(n)$ the execution time of \textit{NeatSort} on a list with $n$ elements and extra space required by it, respecetively, we have
\begin{itemize}
  \item $T(n) = \Omega(n)$ and $T(n) = \mathcal{O}(n\log n)$;
  \item $S(n) = \mathcal{O}(n)$.
\end{itemize}

\section{Disorder metrics}

In this section we will review some of the most common measure of disorder for sorting algorithms and then analyze \textit{NeatSort} performance with respect to them.

The disorder of a sequence is evaluated by a measure of \emph{presortedness} (or measure of disorder), namely a real-valued function over the collection of finite sequences of integers.
More precisely, given a sequence $X$ of distinct elements\footnote{Every sequence with repetitions can be easily mapped to the sequence of unique tuples $(x_i, i)$, where $x_i = X[i]$.}, a measure of disorder $M$ satisfies the following properties:
\begin{enumerate}[label=(\alph*)]
   \item If $X$ is sorted (i.e., if the elements in $X$ are in nondecreasing order, then $M(X)$ = 0.
   \item If $X$ and $Y$ are order isomorphic, then $M(X)$ = $M(Y)$.
   \item If $X$ is a subset of $Y$, then $M(X) \leq M(Y)$.
   \item If every element of $X$ is smaller than every element of $Y$, then $M(X.Y) \leq M(X) + M(Y)$.
   \item $M(\{x\}.X) \leq |X| +  M(X)$, for every $x \in \mathbb{N}$.
\end{enumerate}

The measure of efficiency of a sorting algorithm for a given input array $X$, instead, is the number of comparisons it performs while sorting $X$.

A definition of \emph{optimal} (or \emph{maximal}) \emph{adaptivity} is due to Mannila \cite{mannila1985measures}: a sorting algorithm is optimally adaptive with respect to a measure of disorder if it takes a number of comparisons that is within a constant factor of the lower bound.

Let $\mathit{below}(z, n, M)$ be the set of permutations of $n$ distinct integers whose disorder is not larger to $z$, with respect to a disorder measure $M$, i.e.,
\[
\mathit{below}(z, n, M) = \{ Y \in \mathbb{N}^{< \mathbb{N} } | | Y | = n \wedge  M(Y) \leq z \}\,.
\]

It can be shown that the comparison tree for any sequence $Y$ of length $n$ such that $M(Y) \leq z$ has at least $ | \mathit{below}(z, n, M) | $ leaves, and  so its height is $\Omega ( \log | \mathit{below}(z, n, M) | )$. Hence, for an input array $X$, a comparison based algorithm requires $\Omega(|X| + \log | \mathit{below}(z, |X|, M) |)$ comparisons\footnote{Of course at least a linear number of comparisons is required in order to test presortedness.}.

Mannila \cite{mannila1985measures} defines also the notion of optimal adaptivity in the worst case: let $M$ be a measure of disorder and let $S$ be a sorting algorithm which uses $T_{S}(X)$ comparisons on input $X$. We say that $S$ is optimal with respect to $M$ (or $M$-optimal) if, for some $c > 0$, we have 
\[
T_{S}(X) \leq c \cdot \max \{|X|, \log | \mathit{below}(z, |X|, M) | \}\,,
\]
for every finite sequence $X$ of integers.

\subsection{Commonly used metrics}\label{commonlyUsedMetrics}

In this section we review 11 commonly used measures of disorder.

\begin{enumerate}
  \item \textbf{Inv}: given a sequence $S = \langle s_{1}, s_{2}, \ldots, s_{n}\rangle $, an inversion is any pair $(s_{i}, s_{j})$ such that $i < j$ and $s_{i} > s_{j}$; $Inv(S)$ is the number of inversions in $S$.

  \item \textbf{Dis}: the largest distance determined by an inversion \cite{estivill1989new}. For example, let $S_{1} = \langle 1, 8, 4, 3, 7, 6, 2, 5, 10\rangle $; then $(8, 5)$ is the inversion whose elements are farthest apart, so that $Dis(S_{1}) = 7$. This measure puts more emphasis on the inversions  which are more far apart.

  \item \textbf{Max}: the largest distance an element must travel to reach its sorted position. Let $S_{1}$ as above. Then $8$ must travel 6 positions to reach the right place, so $Max(S_{1}) = 6$. This measure gives more importance to global disorder rather than local disorder.

  \item \textbf{Exc}: the minimum number of exchanges required to sort a sequence \cite{mannila1985measures}. Consider again the sequence $S_{1}$ above. It can be shown that 4 exchanges suffice to sort it, whereas 3 exchanges are not enough. Therefore, $Exc(S_{1}) = 4$.

  \item \textbf{Rem}: the minimum number of elements that must be removed to obtain a sorted subsequence \cite{knuthart}. Considering again our sequence $S_{1}$, we have easily $Rem(S_{1}) = 5$.

  \item \textbf{Runs}: ascending runs are sorted portions of the input; 
for a sequence $S$, $\mathit{Runs}(S)$ is the number of boundaries between the maximal runs in $S$, called step-downs \cite{knuthart2}. Thus, for our example, we have $Runs(S_{1}) = 4$.

\item \textbf{SUS} (short for \textbf{Shuffled Up-Sequences} \cite{levcopoulos1990sorting}); it is a generalization of the \textbf{Runs} measure and is defined as the minimum number of ascending subsequences (of possibly not adjacent elements) into which we can partition a given sequence. In our example, $\mathit{SUS}(S_{1}) = 4$.

\item \textbf{SMS} (short for \textbf{Shuffled Monotone Subsequence}); it further generalizes the previous measure: it is defined as the minimum number of monotone (ascending or descending) subsequences into which one can partition the input sequence \cite{levcopoulos1990sorting}. In our example, $ \mathit{SMS}(S_{1}) = 3 $.

\item \textbf{Enc}: it refers to the concept of \textbf{Encroaching lists} introduced by \textit{Skiena} in its adaptive algorithm \textbf{Melsort} \cite{skiena1988encroaching}; it is defined as the number of sorted lists constructed by \textbf{Melsort} when applied to a sequence.

  \item \textbf{Osc}: it has been defined by \textit{Levcopoulos} and \textit{Petersson} \cite{levcopoulos1989note} after a study of \textbf{Heapsort}; in some sense it evaluates the ``oscillations" of large and small elements in a given sequence.

  \item \textbf{Reg}: this measure has been defined by \textit{Moffat} and \textit{Petersson} \cite{moffat1991historical,petersson1995framework}; it results that any Reg-optimal sorting algorithm is optimally adaptive with respect to the other 10 measures.
\end{enumerate}

A partial order and related equivalence relation on the above measures is provided by the following definition.
\begin{definition}
Let $M_{1}$, $M_{2}$ be two measures of disorder. We state that:
\begin{enumerate}
	\item $M_{1}$ is \emph{algorithmically finer} than $M_{2}$ (denoted $M_{1} \leq_{alg} M_{2}$) if and only if any $M_{1}$-optimal algorithm is also $M_{2}$-optimal.
	\item $M_{1}$ and  $M_{2}$ are \emph{algorithmically equivalent} (denoted $M_{1} =_{alg} M_{2}$) if and only if  $M_{1} \leq_{alg} M_{2}$ and $M_{2} \leq_{alg} M_{1}$.
\end{enumerate}
\end{definition}

\begin{figure}
\centerline{\includegraphics[width=50mm]{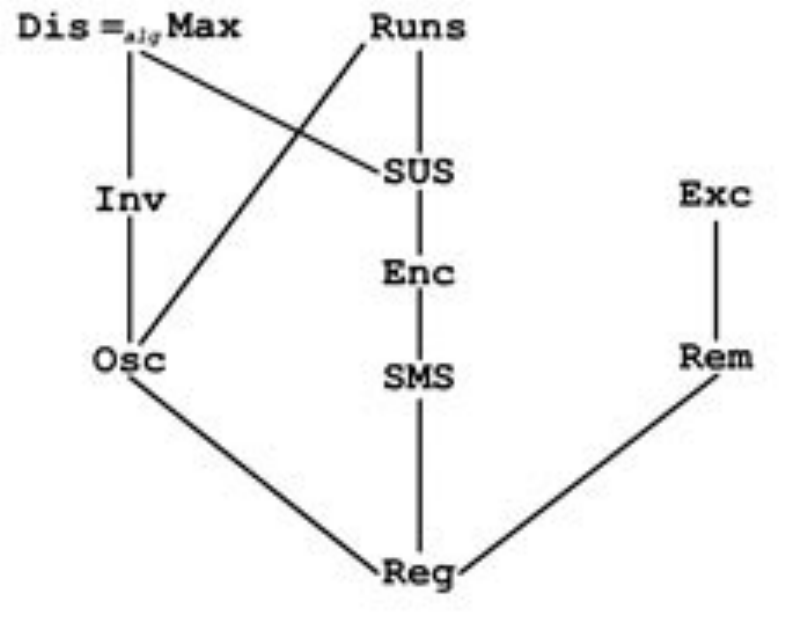}}
\caption{Metrics partial order}
\label{fig:three}
\end{figure}

Figure~\ref{fig:three} shows in details the partial order introduced by $\leq_{alg}$; as already remarked, $\mathit{Reg}$-optimality implies optimality with respect to any other of the above metrics, and  $\mathit{SUS}$-optimality implies  $\mathit{Runs}$-optimality, while it is implied by  $\mathit{SMS}$-optimality.

Therefore, to prove that \textit{NeatSort} is optimal for all these metrics, it is enough to show that it is $Reg$-optimal.

\subsubsection{$\mathit{Reg}$-optimality of \textit{NeatSort}}
$\phantom{empty}$

\noindent Moffat and Petersson \cite{moffat1991historical} defined the measure $\mathit{Reg}$ while studying more efficient variants of \textbf{Insertion-Sort}  which improve  the performance of \textbf{Insertion-Sort} by keeping track of the information gathered during the algorithm execution, such as the position at which the last elements\footnote{The algorithm \textit{Regional Insertion Sort} searches, at each step, a logarithmic fraction of the element in the ordered portion of the array.} have been inserted. 
Let  
\[
d_{i} = \big| \{ k | 1 \leq k < i \wedge \min \{x_{i-1}, x_{i}\} < x_k < \max  \{x_{i-1}, x_{i}\} \}  \big| + 1 
\]
be the distance between the last insertion point to the actual insertion point and let
\[
d_{i, j} = \big| \{ k | 1 \leq k < i \wedge \min \{x_{i}, x_{j} \} < x_k < \max \{x_{i}, x_{j} \}\} \big| + 1
\]
be the distance from $x_{j}$, with $j < i$, to the insertion point of $x_{i}$.  
Note that $d_{i} = d_{i,i-1}$.

Next, for $ i > 1$, let $t_{i} = \min \{ j | 1 < j < i \wedge d_{i, i-j} = 1\} $; $t_{i}$ represents the amount of history needed for inserting $x_{i}$ in its final position.

Finally, by putting $r_{i} = \min \{t+d_{i}, i-t \}$, we then have  $\mathit{Reg}(X) = \prod_{i=2}^{|X|}(r_{i}-1)$.

Since every sublist $L[q]$ is ordered after \textit{NeatSort}'s preliminary phase, $r_{i}=1$ for $i=1,..,|X|$, and therefore $\sum_{q=0}^{m}Reg(L[q]) = 0$
and thus \textit{NeatSort} is adaptive with respect to the measure $Reg$, and it is also optimal for all the other measures defined above.

\subsubsection{Metrics Lower Bounds for \textit{NeatSort}}
\label{sec:neatsort_metrics}
$\phantom{empty}$

\noindent 
Estivill-Castro and Wood introduced, in 1992 \cite{estivill1992survey}, the notion of \emph{generic sorting algorithm} (see Algorithm~\ref{alg:three} below).

\begin{algorithm}[H]
\SetAlgoNoLine
\KwIn{A list $X$ with $n$ elements.}
\KwOut{The ordered version of the input list.}

\If{$X$ is sorted}{terminate;}
\If {$X$ is simple}{sort $X$ using an alternative sorting algorithm for simple sequences;}
\ElseIf{$X$ is neither sorted nor simple}{
	apply a division protocol to divide $X$ into at least $s \geq 2$ disjoint sequences;
	
	recursively sort the sequences using Generic Sort;

	merge the sorted sequences to obtain $X$ in sorted order;
}
\caption{Generic Sort}
\label{alg:three}
\end{algorithm}
\begin{remark}
The definition of ``simple" in Algorithm~\ref{alg:three} depends on the actual definition of the algorithm.
\end{remark}

\begin{center}%
\captionof{table}{Known lower bounds for disorder metrics}
\begin{tabular}{|c|c|}
\hline
\textbf{Measure}	& \textbf{Lower bound: $ \log \| below(M(X),|X|,M) \| $ } \\\hline
$Dis$ & 	$\Omega(|X|(1 + \log(Dis(X)+1) ))$	\\\hline
$Exc$ & 	$\Omega(|X|(1 + Exc(X)\log(Exc(X)+1) ))$	\\\hline
$Enc$ & 	$\Omega(|X|(1 + \log (Enc(X) + 1) ))$	\\\hline
$Inv$ & 	$\Omega(|X| \cdot (1 + \log(\frac{Inv(X)}{|X|} +1 ) ))$	\\\hline
$Max$ & 	$\Omega(|X|(1 + \log(Max(X)+1) ))$	\\\hline
$Osc$ & 	$\Omega(|X| \cdot (1 + \log(\frac{Osc(X)}{|X|} +1 ) ))$	\\\hline
$Reg$ & 	$\Omega(|X|(1 + \log(Reg(X)+1) ))$	\\\hline
$Rem$ & 	$\Omega(|X|(1 + Rem(X)\log(Rem(X)+1) ))$	\\\hline
$Runs$ & 	$\Omega(|X|(1 + \log(Runs(X)+1) ))$	\\\hline
$SMS$ & 	$\Omega(|X|(1 + \log(SMS(X)+1) ))$	\\\hline
$SUS$ & 	$\Omega(|X|(1 + \log(SUS(X)+1) ))$	\\\hline

\end{tabular}
\label{tab:one}
\end{center}

As is clear, \textit{NeatSort} perfectly fits the description above. We can thus make use of the following theorem \cite{estivill1990generic}:
\begin{theorem}
Let $M$ be a measure of disorder such that a sequence $X$ is simple whenever
$M(X) = 0$, and let $D \in \mathbb{R}$ and $s \in \mathbb{N}$ be constants such that $0 \leq D < 2$ and $s > 1$. Also, let $\mathit{DP}$ be a linear-time division protocol that divides any sequence $X$ into $s$ sequences of almost equal sizes.
Then:
\begin{enumerate}
  \item \textit{Generic Sort} is worst-case optimal and it takes
  $\mathcal{O}\big(\left|X\right|\log\left|X\right|\big)$-time in the worst
  case.
  \item 
  Generic Sort is adaptive with respect to the measure $M$ and it takes $\mathcal{O}\big(\left|X\right|\cdot ( 1+\log (M(X)+1) ) \big)$-time in the worst case, provided that 
  \[
  \sum_{j=1}^{s}M\left(\text{$j$-{th} sequence}\right)\leq D\cdot\left\lfloor \frac{s}{2}\right\rfloor \cdot M\left(Y\right)
  \]
  holds, for all sufficiently long sequences $Y$.
%
\end{enumerate}
\end{theorem}

Table~\ref{tab:one} reports the known lower bounds for the metrics defined in Section~\ref{commonlyUsedMetrics}: \textit{NeatSort}, as proved above, being optimal for all these metrics, meets all such lower bounds.

\section{Performance}

\begin{algorithm}[H]
\SetAlgoNoLine
\KwIn{A list $X$ of length $n$.}
\KwOut{The ordered version of the input list.}

$listCount$ := $1$;

\textbf{put} $X_1$ in $list_1$;

\For{$i$ := $2$ to $n$}{
  \For{$j$ := $1$ to $\mathit{listCount}$}{
    \If{$X_i < head(list_j)$}{
      \textbf{add} $X_i$ to the head of $list_j$;
      
      \textbf{break};
    }
    \ElseIf{$X_i > tail(list_j)$}{
      \textbf{add} $X_i$ to the tail of $list_j$;
      
      \textbf{break};
    }    
  }
  \If{$X_i$ couldn't be added to any list}{
    \textbf{add} $1$ to $\mathit{listCount}$;
    
    \textbf{create} $\mathit{list}_{\mathit{listCount}}$;
    
    \textbf{put} $X_i$ in the newly created list;
  }
}

\While{$\mathit{listCount} > 1$}{
  \If{$\mathit{odd}(\mathit{listCount})$}{
    $head(\mathit{listCount}-1) := \textbf{merge}(head(\mathit{listCount}-1), head(\mathit{listCount}))$;
  }
  \For{$i$ := $1$ to $|\frac{listCount}{2}|$}{
		$head(i):= \textbf{merge}\Big(head(i), head(|\frac{\mathit{listCount}}{2}|+i)\Big)$;
	}
  $\mathit{listCount}$ /:= 2;
}

\Return head(1)

\caption{Melsort}
\label{alg:four}
\end{algorithm}

To test the performance of our algorithm, a test suite has been designed to benchmark \textit{NeatSort} behavior against a tuned version of random \textit{Quicksort} algorithm, the standard C++ \textbf{qsort} function, a tuned version of \textit{Mergesort}, and \textit{Skiena's} \textbf{Melsort} algorithm (whose pseudo code is shown in the box for Algorithm~\ref{alg:four}).

To minimize the influence of kernel and other background programs running at the same time, the test suite iterates a loop executing in turn all 5 algorithms, once per iteration, on (a copy of) the same array; these arrays are generated randomly (or according to specific criteria) at every iteration. 
In this way, possible computational lags due to external factors will affect on average all the algorithms much in the same way.

The test suite has been run on different machines:
\begin{itemize}
\item a desktop PC with an Intel core-duo processor and 2 GB of RAM, running Windows Vista, 32 bit version;

\item an Asus notebook with an Intel Core i7 2.0 GHz processor, 6 GB of RAM and running both Windows 7, 64 bit version and, in a separate partition, Ubuntu 10, 64 bit version;

\item a Fujitsu Siemens notebook with an Intel core-duo P8400 processor (2.26 GHz), 4 GB of RAM and running Ubuntu 10, 64 bit version.
\end{itemize}

Under Windows, Microsoft Visual C++ Express has been used, setting the compiler to make advantage of the multicore processor and to optimize the code for faster execution. 
Under Linux, the Netbeans 6.9.1 suite had been used with the g++ compiler set for multicore processor 64 bit machines.

The simulation has provided consistent results on all the platforms.
To ensure the greatest precision in evaluating algorithms' performance, it has been used the high resolution time measure mechanism provided by both systems: by window.h library in Windows (the minimum measurable interval is approximately 10 microseconds, with a resolution of 1190000 tick per second) and by time.h library in Linux; using the \textit{clock-gettime} function, the interval resolution is 1 nanosecond.
At each iteration, for every algorithm the number of intervals consumed is stored and then, at the end of the cycle, the median value is extracted for each algorithm; the median value, unlike the average (that is computed anyway), is not affected by extreme, out of scale, values, which can be caused by unpredictable peaks of requests for OS' services: this is especially true for large testing sets.
Both values (median and average) are expressed in milliseconds and rounded to the microsecond.

The first test suite has been run on random arrays, then a few specific cases are examined: ordered arrays, inversely ordered arrays, and partially ordered ones.
We tried to make as an extensive test as it was possible, considering the time requested to sort huge arrays; in details, the number of iteration has been fixed depending on array's size.

\begin{center}%
\captionof{table}{Relation between array size and number of test cases}
\begin{tabular}{|l|l|}
\hline
Array size & Test cases\\\hline
3.276.800 & 5000\\\hline
Da 100 a 102.400 & 10000\\\hline
Da 204.800 a 409.600 & 50000\\\hline
819.200 & 25000\\\hline
1.638.400 & 10000\\\hline
Da 6.553.600 a 26.214.400 & 1000\\\hline
50.000.000+ & 500\\\hline
\end{tabular}
\label{tab:two}
\end{center}%



\subsection{Random Arrays}
\label{sec:randarrays}

Tests on random arrays show consistent performance for \textit{NeatSort} as the size of the arrays grow. For small arrays, the best performing 
algorithm is the implementation of random \textit{Quicksort}, provided \href{https://gist.github.com/mlarocca/fa340347b63c70ce56c2}{here}, that had been optimized for best performance. 
For larger arrays, however, this algorithm's performance 
progressively degrades, while \textit{NeatSort} and \textbf{qsort} 
steadily grows with $n\log n$, as highlighted using a logarithmic scale to visualize the results (Figure~\ref{fig:six}).
The results have been averaged over all the testing platforms.

\begin{center}

\centerline{\includegraphics[width=150mm]{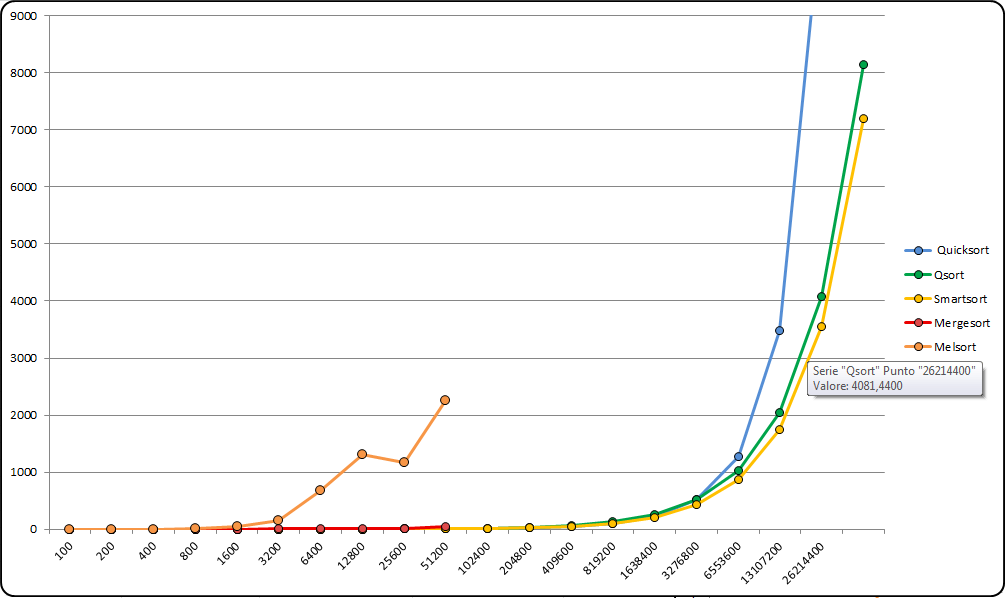}}
\captionof{figure}{Execution time (ms) on random arrays}
\label{fig:four}
\end{center}

\begin{center}

\centerline{\includegraphics[width=150mm]{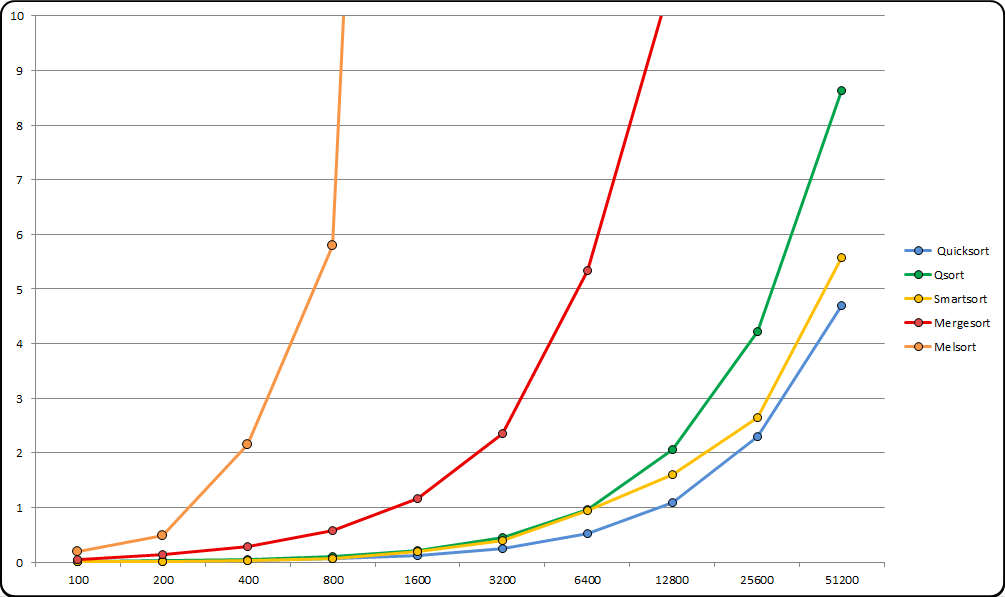}}
\captionof{figure}{Details of previous chart for arrays of size $\leq 50K$ elements}
\label{fig:five}
\end{center}

\begin{center}

\centerline{\includegraphics[width=150mm]{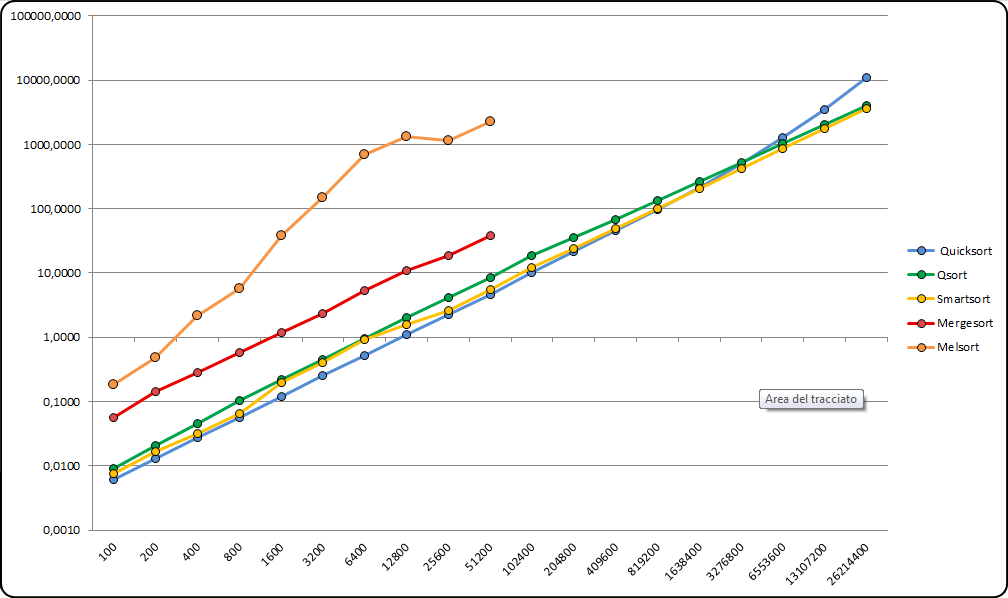}}
\captionof{figure}{Execution time (ms) on random arrays - $\log_{10}$ scale}
\label{fig:six}
\end{center}

\subsection{Sorted arrays (Most favourable case)}
\label{sec:sortarrays}

Sorted arrays are the most favourable case for \textit{NeatSort}, and indeed the measured performance shows that \textit{NeatSort}'s running time is several order of magnitudes smaller than the other algorithms.

\begin{center}
\centerline{\includegraphics[width=150mm]{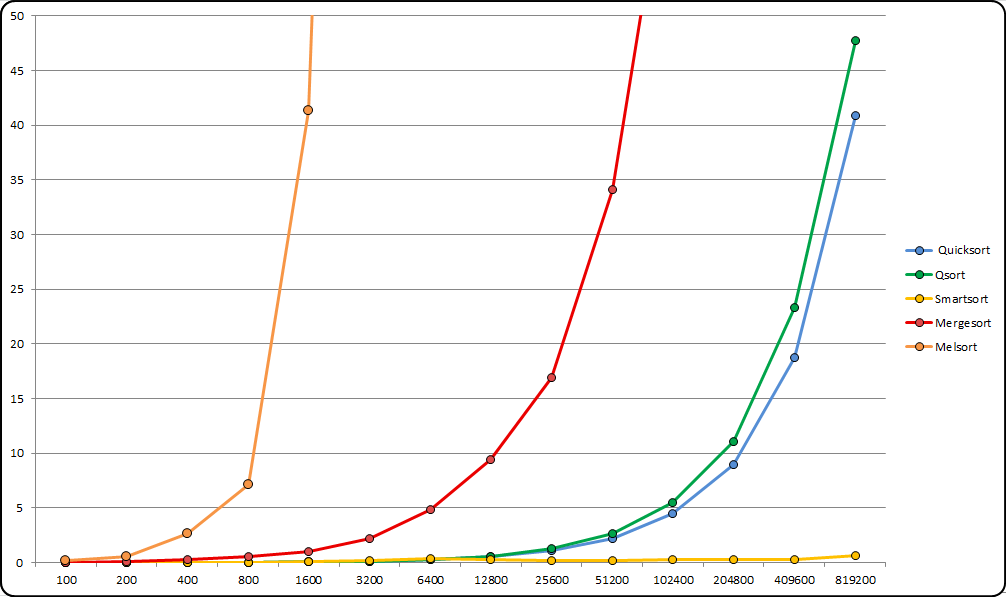}}
\captionof{figure}{Execution time (ms) on sorted arrays}
\label{fig:seven}
\end{center}

\begin{center}
\centerline{\includegraphics[width=150mm]{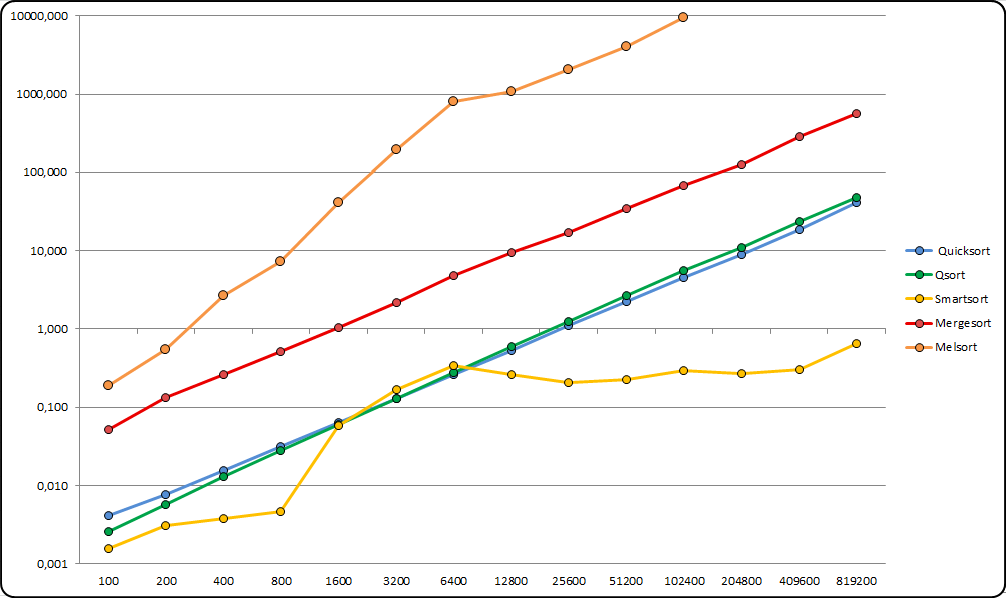}}
\captionof{figure}{Execution time (ms) on sorted arrays - $\log_{10}$ scale}
\label{fig:eight}
\end{center}

\subsection{Statistics about performance and disorder metrics}

So far, we have only examined the two extremes of the input 
landscape; sorted arrays are, by design, the most favourable case for 
\textit{NeatSort}, but it would be reasonable to expect that when run on nearly sorted arrays the algorithm would largely benefit from the analysis phase and demonstrate superior performance.
To further investigate this issue, we run a series of comparative 
tests on \textbf{qsort} and \textit{NeatSort}, gathering, together with performance measurements, a set of statistics about the degree of disorder of the input, with the goal of bringing up correlations between the relative performance, and \begin{itemize}
  \item the number of \textit{inversions},
  \item the \textit{max distance} of elements to their position in the sorted sequence,
  \item the number \textit{runs} in the input array.
\end{itemize}

For each of these metrics, two charts are shown:
\begin{enumerate}
  \item a 2D chart stressing correlation between the metric and the relative performance of \textit{NeatSort} in comparison to \textbf{qsort};
  \item a 3D chart, where each point in the $\mathbb{R}^2$ domain correspond to an input sequence identified by its size and the measure for the metric.
\end{enumerate}

In both charts, the relative performance is expressed in percentage, and computed as
\[
\frac{T_{\mathit{qsort}} - T_{\mathit{NeatSort}}}{\max(T_{\mathit{qsort}}, T_{\mathit{NeatSort}})}\times 100\,.
\] 
So positive values show better performance for \textit{NeatSort} 
(the greater the absolute value, the better), and negative values, instead, shows cases in which \textbf{qsort} outperformed \textit{NeatSort}. Values are shown using a gradient going from green (for positive values), to yellow (for ties), to red (for negative values).

In the 2D charts, the size of the dots is proportional to the size of the test case.

\subsubsection{Inversions}
\label{sec:inversions}

The number of inversions is shown as a percentage of the maximum number of possible inversions for the input size: for a sequence of length $n$, there can be at most $\frac{n  (n-1)}{2}$ inversions.

As expected, the data plots a bowl-shaped figure with a minimum corresponding to $50$\% of inversions, while ordered sequences ($0$\% of inversions) and reversed sequences ($100$\% of inversions) represent the best case scenario for \textit{NeatSort}.
The figure also shows a different cluster, showing almost constant relative performance, in correspondence with larger input sequences. As it is also clarified by the 3D chart, this anomaly in the results testify that the performance delta in favour of \textit{NeatSort} grows with the size of the input.

\begin{center}
\centerline{\includegraphics[width=150mm]{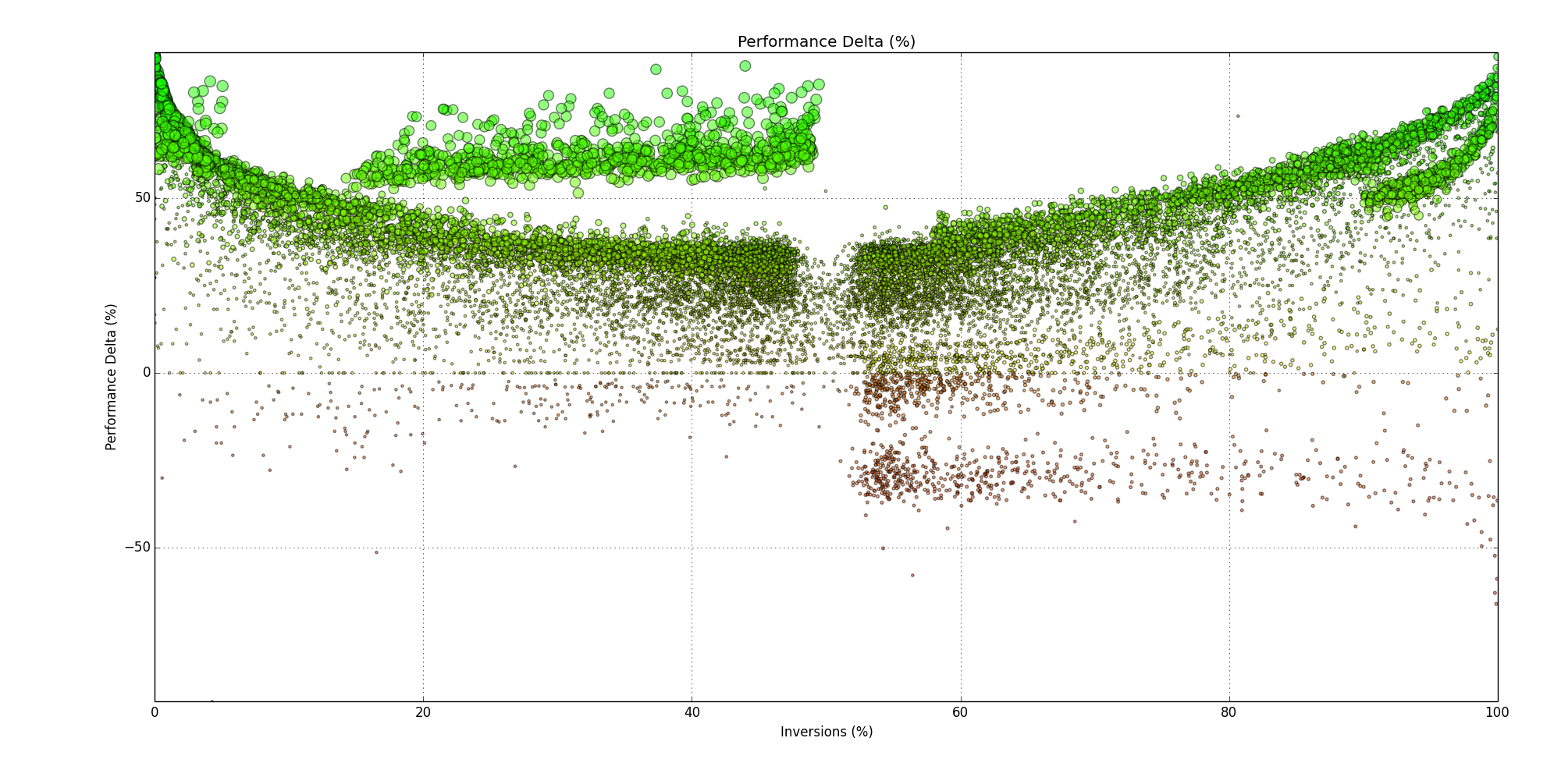}}
\captionof{figure}{Relative performance for \textit{NeatSort} and \textit{qsort}, with respect to percentage of inversions (spots proportional to array size)}
\label{fig:nine}
\end{center}

\begin{center}
\centerline{\includegraphics[width=150mm]{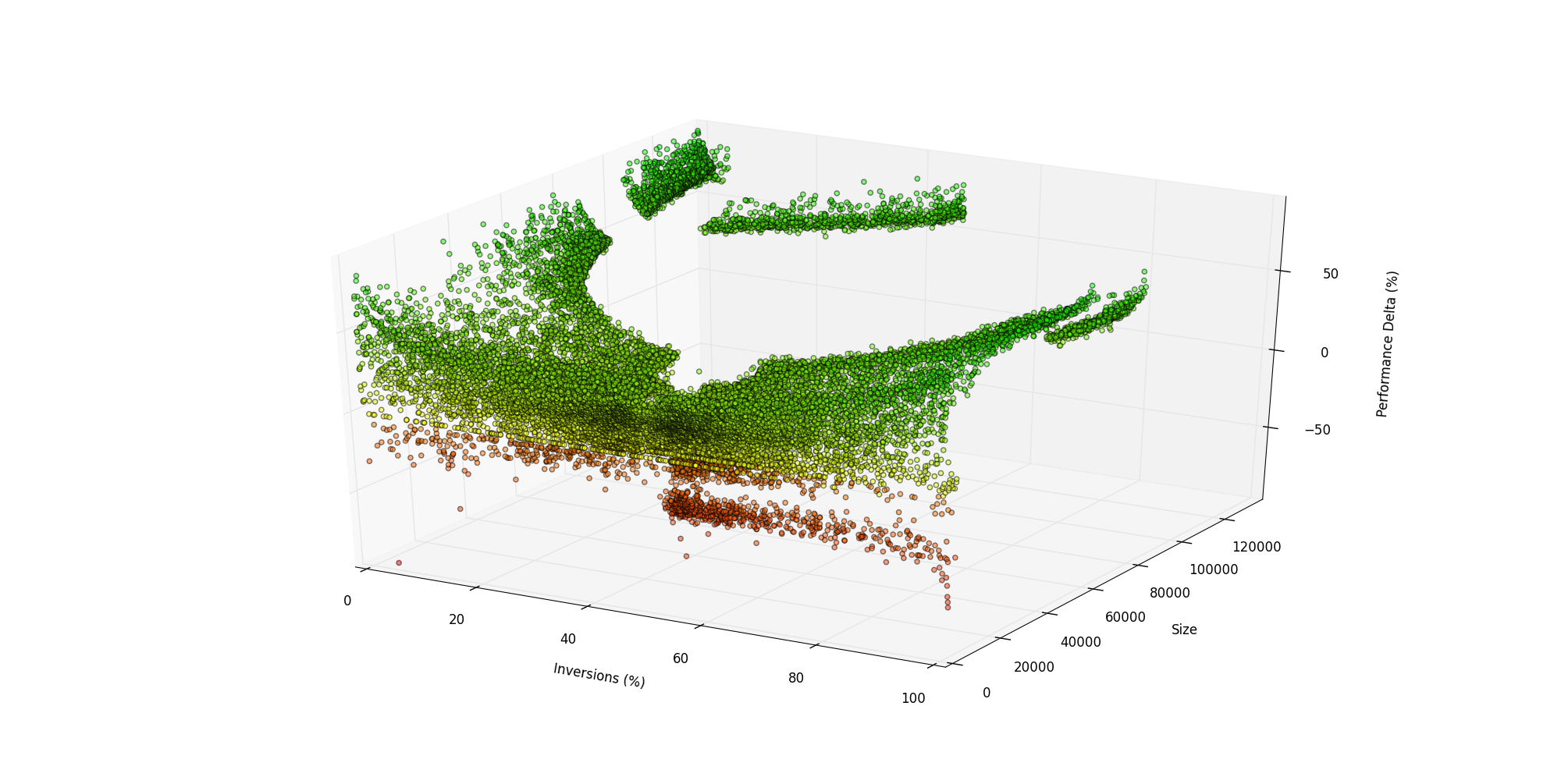}}
\captionof{figure}{Relative performance for \textit{NeatSort} and \textit{qsort}, with respect to percentage of inversions and size}
\label{fig:ten}
\end{center}

\subsubsection{Max distance}

The charts in this section show the relative performance with respect to the maximum distance of elements in the input (expressed as a percentage of the input length).
Interestingly enough, \textit{NeatSort}'s relative performance steadily improves not only as the max distance becomes smaller, but also as the size of the input grows.

\begin{center}
\centerline{\includegraphics[width=150mm]{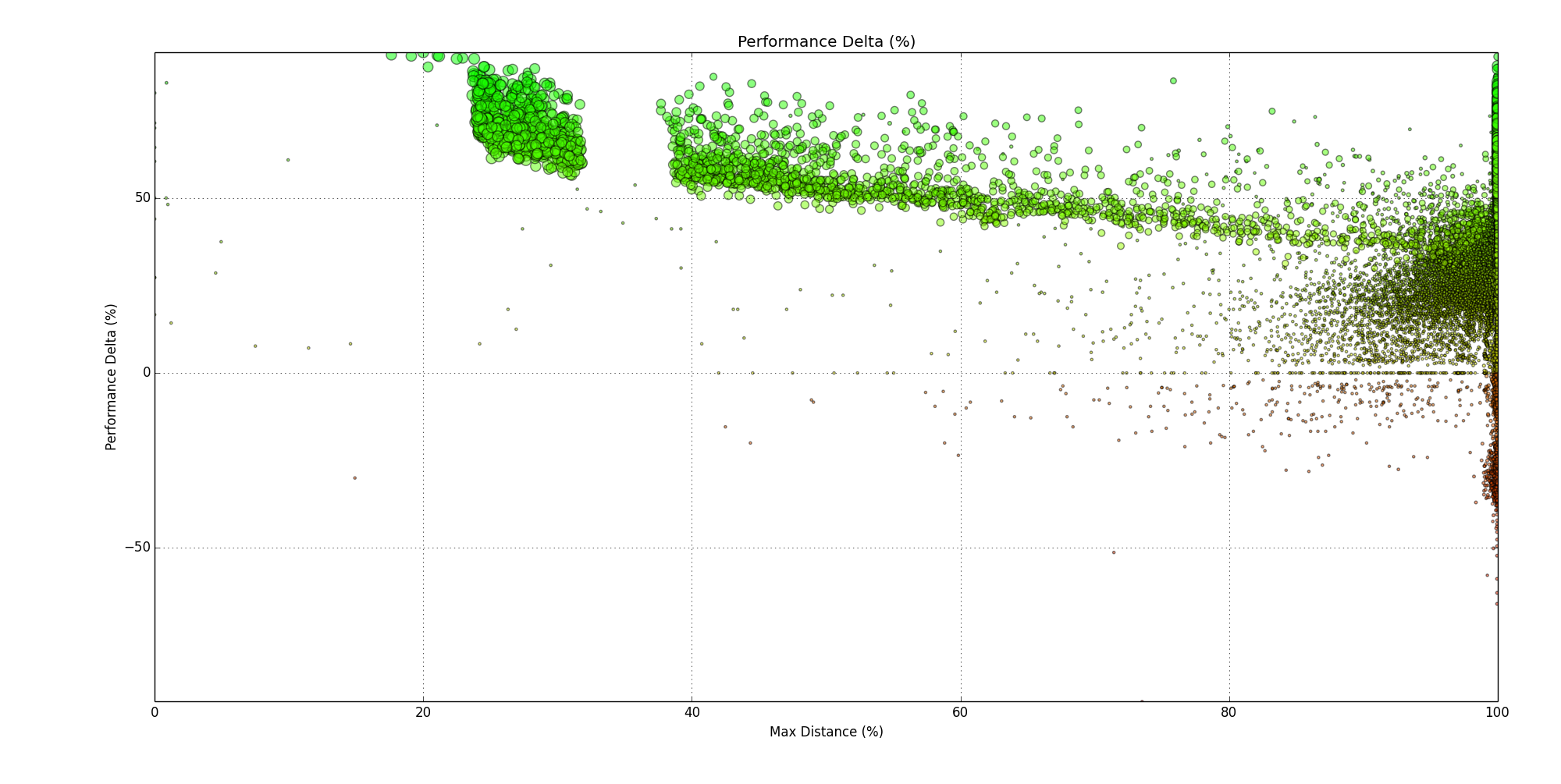}}
\captionof{figure}{Relative performance for \textit{NeatSort} and \textit{qsort}, with respect to \textit{max distance / array size} (spots proportional to array size)}
\label{fig:eleven}
\end{center}

\begin{center}
\centerline{\includegraphics[width=150mm]{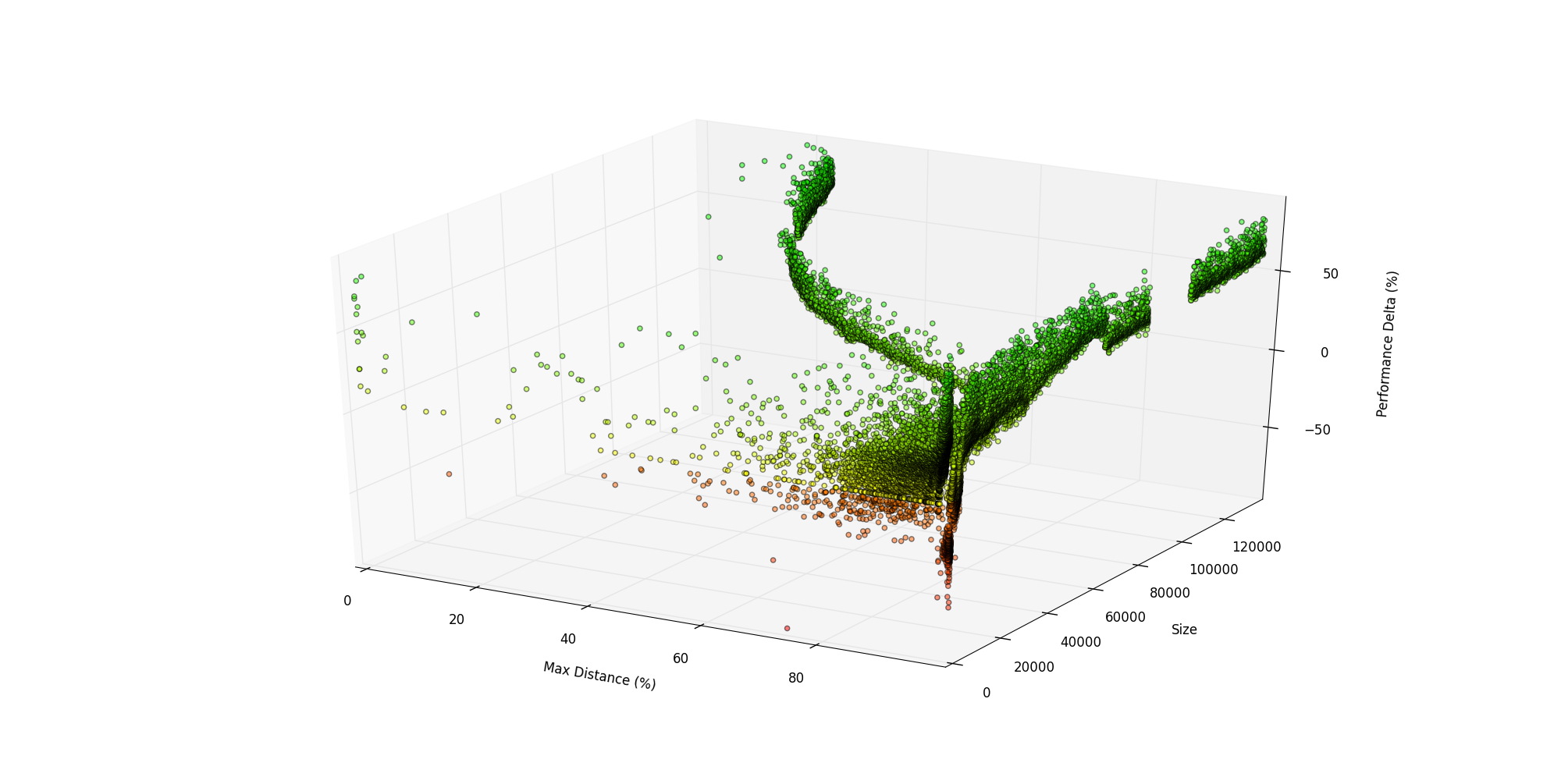}}
\captionof{figure}{Relative performance for \textit{NeatSort} and \textit{qsort}, with respect to \textit{max distance / array size} and size}
\label{fig:twelve}
\end{center}

\subsubsection{Runs}

The number of \textit{runs} is expressed as a percentage of the input length; Figures~\ref{fig:thirteen} and~\ref{fig:fourteen} show, as expected, much the same shape as Figure~\ref{fig:nine}: $0$\% \textit{runs} corresponds to sorted sequence, but as \textit{runs} grows from $50$\% (the global minimum for relative performance) to $100$\% (and hence toward reversed sequences), \textit{NeatSort} performs increasingly better.

Interestingly, a local maximum is present in correspondence to sequences with nearly $60K$ elements and $78$\% of runs.

\begin{center}
\centerline{\includegraphics[width=150mm]{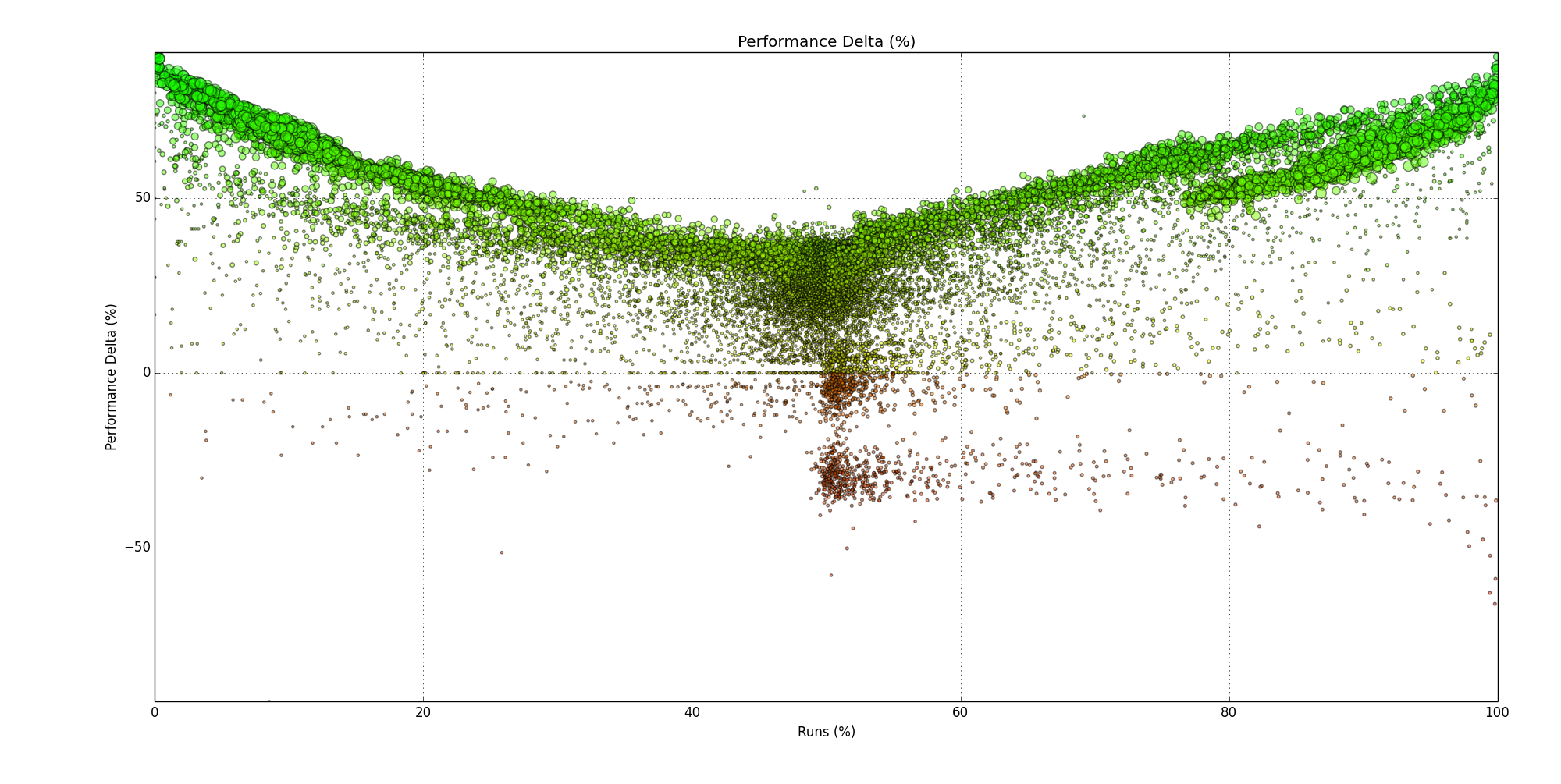}}
\captionof{figure}{Relative performance for \textit{NeatSort} and \textit{qsort}, with respect to runs/size (spots proportional to array size)}
\label{fig:thirteen}
\end{center}

\begin{center}
\centerline{\includegraphics[width=150mm]{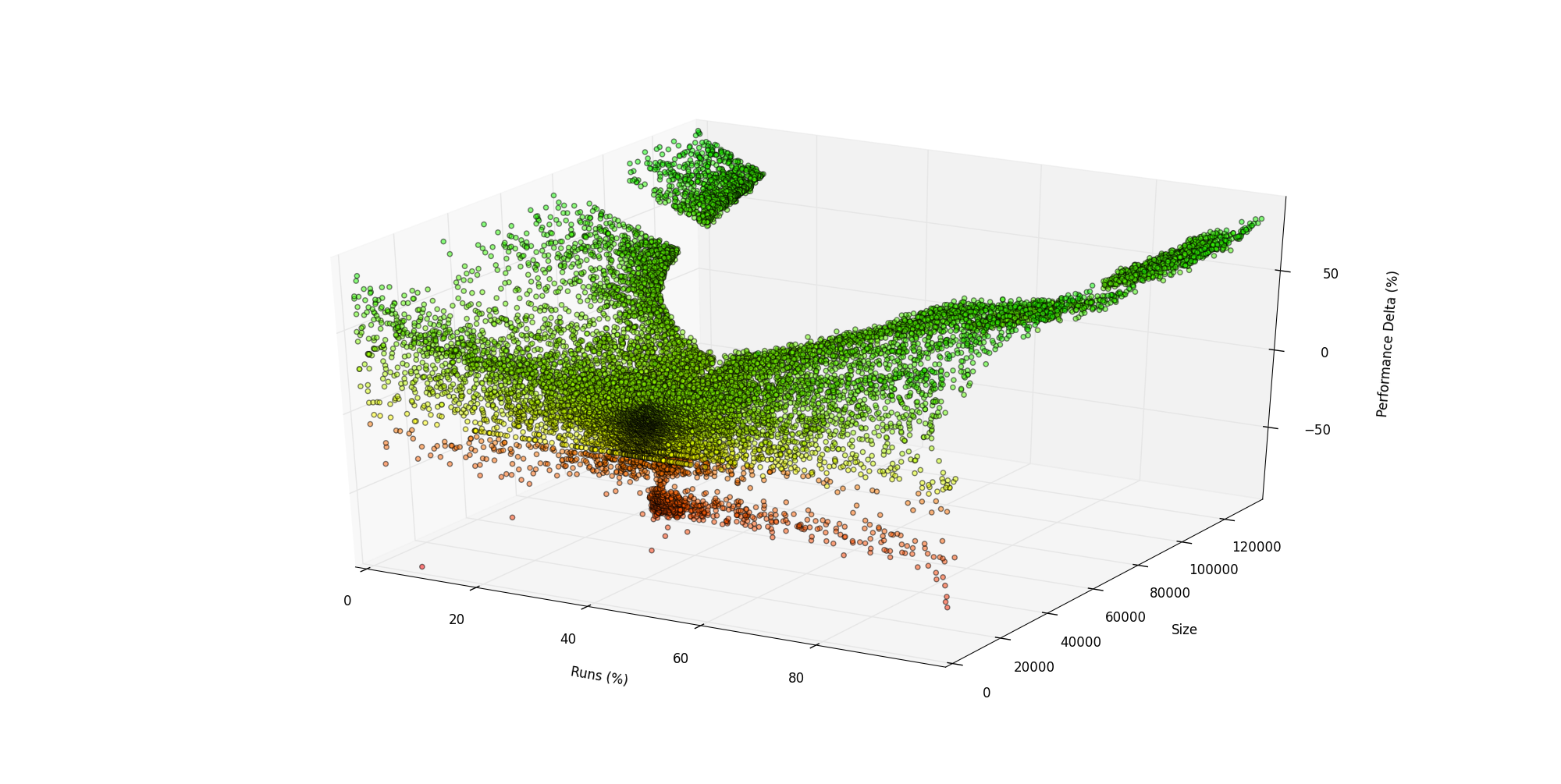}}
\captionof{figure}{Relative performance for \textit{NeatSort} and \textit{qsort}, with respect to runs/size and size}
\label{fig:fourteen}
\end{center}

\section{Conclusions}

We have presented an intuitive adaptive sorting algorithm that proves to be optimal for most of the disorder metrics present in literature.
Although other algorithms exists that outperforms it on peculiar ad hoc metrics (in particular \textit{Melsort} \cite{skiena1988encroaching}), those algorithms, as most of the adaptive algorithms in literature, have such an intricate workflow that their implementations results slower by some order of magnitude in comparison with \textit{Mergesort} or \textit{Quicksort}.
For \textit{NeatSort}, instead, we have carefully both engineered its design to be as simple as possible and tuned its implementation to make it extremely efficient and performant.
The result is a flexible and fast algorithm which on average is as efficient as \textit{Quicksort} and outperforms even the \textit{C} implementation of \textit{qsort}: the ratio between the running times of \textit{Neatsort} and \textit{qsort}, besides being consistently below 1, gets progressively smaller as the number of inversions moves from 50\% to both 0\% and 100\%, i.e. to sorted sequences in direct and inverse order.



\section{Acknowledgments}
  Charts in Sections~\ref{sec:randarrays} and~\ref{sec:sortarrays} have been created with Excel\textcopyright Starter 2010, while the remaining charts have been created with the \href{http://matplotlib.org/}{MatPlotLib} Python library. 


\bibliographystyle{plainnat}
\addcontentsline{toc}{section}{\refname}\bibliography{neatsort}


\medskip

\end{document}